\documentstyle[aps,preprint,epsf]{revtex}

\begin{document}
\preprint{\sl Submitted to Phys. Rev. Letters}

\title{On thermoelastic phenomena around orientational \\ ordering
        transition in crystalline C$_{60}$}

\author {M.A.Fradkin\cite{IKAN}}

\address{Department of Mechanical and Aerospace Engineering  \\
                Carleton University, Ottawa, Ont. K1S 5B6, Canada \\
                {\em e-mail:} {\tt mfradkin@next.mrco.carleton.ca}}

\draft
\maketitle

\begin{abstract}
Thermoelastic phenomena associated with orientational phase transition
in solid C$_{60}$ are considered. Coupling of the order parameter with
elastic strain is analyzed in the Landau theory of phase transitions.
Wide range of possible coexistence of the FCC phase with SC one in
C$_{60}$ has been found from 260 K down to critical temperature 160 K
of the instability of high-symmetry phase. The specific heat, thermal
expansion coefficient and isothermal compressibility for low-symmetry
phase are calculated as function of temperature and pressure in an
agreement with reported data. Applicability of the Landau theory to
ordering transition in fullerenes is discussed.
\end{abstract}

\pacs{05.70.Fh, 64.10.+h, 64.70.-p}

\narrowtext

Among other crystalline fullerenes solid C$_{60}$ exhibits
orientational ordering transition \cite{HeineyFisher}. There are
extensive experimental data available about crystallography of this
transition, associated anomalies of bulk modulus and thermal expansion
coefficient as well as effect of applied external pressure, however,
no adequate theory of such thermoelastic phenomena has been developed
so far. Phenomenological approach based on the mean-field Landau
theory of ordering transition is used in the present paper to analyze
these effects. Due to high symmetry of the buckyball C$_{60}$
molecules microscopical theory of ordering transition involves
symmetry adapted rotator functions to describe orientational
interaction and fluctuations \cite{MCN}. However, spontaneous lattice
distortion originates from coupling of the orientational order
parameter with homogeneous strain and, hence, mean-field Landau theory
\cite{MichelLamoen,Harris} can be employed.

Thermodynamic potential around the symmetry breaking phase transition
can be expanded in this theory \cite{LandLif} in powers of 'order
parameter' $\eta$. The theory initially was developed for second-order
transition where order parameter is continuous at the transition
point, however weakly-discontinuous first-order transitions like the
orientational ordering transition in solid C$_{60}$ can be considered
as well \cite{Toledano}. If the symmetry groups of both parent and
product phases are known {\em a priory} then scalar order parameter
can be used and the Ginzburg-Landau expansion of the free energy
difference between the parent and product phases has the form
\cite{LandLif,Toledano}
\begin{equation}
\Delta {\cal G} = {\alpha \over 2}(T - T_c)\eta^2 + {B \over 3}\eta^3  +
	{C \over 4}\eta^4 - E \eta . \label{firstfield}
\end{equation}
where $T_c$ is a critical temperature and $E$ is external conjugated
field. Only second-degree coefficient is supposed to depend on
temperature and equilibrium value of $\eta$ is determined by the
minimization of $\Delta {\cal G}$. The stability requires highest
order coefficient $C$ to be positive and the third degree term $B \neq
0$ implies first order of the transition.

In the $E = 0$ case the Gibbs free energy (\ref{firstfield}) has two
possible minima. One with $\eta = 0$ corresponds to a high-temperature
undistorted phase stable for $T \ge T_c$. For $$T \le T_0 = T_c + {1
\over 4}{B^2 \over {\alpha C}}$$ low-symmetry phase exists with
\begin{equation}
        \eta = - {B \over {2 C}} \left(1 + \left({T_0 - T \over {T_0 - T_c}}
	\right)^{\frac{1}{2}} \,\right) . \label{FOE}
\end{equation}
The phase energies become equal at the temperature of first-order
transition $T_\ast = T_c + {2 \over 9}{B^2 \over {\alpha C}}$ where
$\eta$ jumps to $- {2 \over 3}\,{B \over C}$. The temperature range
between $T_c$ and $T_0$ where both high- and low symmetry phase are
stable and can, thus, co-exist is an absolute thermal hysteresis, real
hysteresis being determined through the nucleation and growth kinetics
by actual cooling/heating rate.

The temperature dependence of specific heat in the low-symmetry phase
\begin{eqnarray}
\Delta {\cal C}_P (T) & = & - T
\frac{\partial^2 \Delta {\cal G}}{\partial T^2} \nonumber \\
& = & {T \over 2}{{\alpha}^2 \over C} \left(1 +
\left({T_0 - T_c \over {T_0 - T}}\right)^{\frac{1}{2}} \,\right)
\label{specheat}
\end{eqnarray}
implies divergence $\propto \ (T_0 - T)^{-\frac{1}{2}}$ as $T
\rightarrow T_0$. This can be seen as some sort of 'precursor'
phenomena {\em below} the transition temperature in a contrast with
the critical region near the continuous transition, where specific
heat anomaly due to critical fluctuations around $T_c$
\begin{equation}
\langle \eta^2 \rangle \simeq {k_B T \over \alpha\,|T - T_c|} \ ,
\label{fluct}
\end{equation}
appears both in high- and in low-symmetry state.

Full review of the Landau theory of ordering transitions in
crystalline fullerenes can be found elsewhere \cite{Harris}. The
transition in C$_{60}$ takes place around 260 K and reduces
crystallographic symmetry from FCC (or $Fm\bar{3}m$) lattice into SC
one with space group $Pa\bar{3}$ \cite{heiney,David91}. The
phenomenological order parameter $\eta$ defined by the symmetry change
belongs to representation of $Fm\bar{3}m$ space group corresponding to
$X$ point of the Brillouin zone \cite{Harris}.

To analyze thermoelastic phenomena the coupling of $\eta$ with
homogeneous strain tensor $\hat{\epsilon}$ should be considered and
the Ginzburg-Landau expansion should include the terms associated with
elastic energy. It was shown \cite{MichelLamoen} that symmetry allows
volumetric strain $\epsilon_0 = {\rm Tr} (\hat{\epsilon})$ to be
coupled with $\eta$ in the lowest order by term
\begin{equation}
        \Delta {\cal G}_{int} = D_0 \epsilon_0 \eta^2 \ .
        \label{inter}
\end{equation}
Any shear strain has a higher degree of coupling with the order
parameter and can be neglected in the present consideration.
Furthermore, there might be no external homogeneous field conjugated
to the order parameter associated with wave vector ${\bf q} \ne 0$.
Coupling between the order parameter and volume change similar to
Eq.(\ref{inter}) has been recently considered \cite{me} in
phenomenological theory of weakly-discontinuous ferroelastic phase
transition in cubic crystals. It was shown that the minimization of
free energy with respect to $\epsilon_0$ leads to renormed
Ginzburg-Landau expansion of $\Delta {\cal G}$ in powers of $\eta$
\cite{me}:
\begin{eqnarray}
\Delta {\cal G}(T,P,\eta) & = & \Delta {\cal G}_0 +
{\alpha^{\prime} \over 2}\,(T - T_c^{\prime})\,\eta^2 \nonumber \\
& & \ + {B \over 3}\, \eta^3 + {C^{\prime} \over 4} \,\eta^4 - E \eta \ ,
\label{RTOT}
\end{eqnarray}
with coefficients
\begin{mathletters}
        \label{coeff}
\begin{eqnarray}
\alpha^{\prime} & = & \alpha + 2 \alpha_{_V} D_0  \\
T_c^{\prime} & = & T_c + {{2 D_0} \over {K_0 \alpha^{\prime}}}\,P
	\label{tempres} \\
C^{\prime} & = & C  - {{2 D_0^2} \over K_0} \ .
\end{eqnarray}
\end{mathletters}
Here $\Delta {\cal G}_0\,(T,P)$ is thermal expansion energy of a
high-symmetry phase with $\eta = 0$ \cite{LanLifEl}
\begin{equation}
\Delta {\cal G}_0(\epsilon_0) = - \alpha_{_V} K_0 (T - T_c)\,\epsilon_0
+ {K_0 \over 2}\epsilon_0^2 + P \epsilon_0 \ , \label{volume}
\end{equation}
with $\alpha_{_V}$ and $K_0$ being the volume thermal expansion
coefficient and bulk modulus of the high-symmetry phase, respectively.
The applied hydrostatic pressure $P$ shifts the transition
temperature:
\begin{equation}
{{dT_c}^{\prime} \over {dP}} = {{2 D_0 } \over {K_0 \alpha^{\prime}}}.
\label{dtdp}
\end{equation}

The expressions for transition-related anomalies in isothermal
compressibility and thermal expansion coefficient can be obtained
\cite{me} by substituting equilibrium $\eta(T,P)$ for low-symmetry
phase into Ginzburg-Landau expansion and then differentiating with
respect to temperature and/or pressure. The difference in thermal
expansion coefficient between low- and high-symmetry phase has the
following dependence on the temperature
\begin{eqnarray}
\label{expcoeff1}
\Delta \alpha_{_V} (T) & = & \frac{\partial^2}{\partial P \partial T}
\left( \Delta {\cal G}- \Delta {\cal G}_0 \right) \nonumber \\
& = & {{\alpha^{\prime} D_0} \over {C^{\prime}} K_0} \left(1 + \left(
{T_0 - T_c \over {T_0^{\prime} - T}} \right)^{\frac{1}{2}}\,\right) \ .
\end{eqnarray}
This expression along with analogous one for compressibility $\Delta
\beta_T$ \cite{me} are similar to Eq.(\ref{specheat}) for $\Delta
{\cal C}_P$ and imply 'precursor'-like divergence near $T_0$. The
discontinuities at the temperature $T_{\ast}^{\prime}$ of the
first-order transition
\begin{equation}
\Delta {\cal C}_P = 2 T_{\ast}^{\prime}
{{{\alpha^{\prime}}^2} \over {C^{\prime}}} , \ \
\Delta \alpha_{_V} = {4 {\alpha^{\prime} D_0} \over {K_0 C^{\prime}}}
, \ \ \Delta \beta_T = {{8 D_0^2} \over {C^{\prime} K_0^2}}
\label{discont}
\end{equation}
are related by the Keesom-Ehrenfest relationships \cite{me}
\begin{equation}
        \label{keeserenf}
{{dT_c^{\prime}} \over {dP}} =
{{\Delta \beta_T} \over {\Delta \alpha_{_V}}} =
{{T_{\ast}^{\prime} \Delta \alpha_{_V}} \over {\Delta {\cal C}_P}}\ .
\end{equation}

Several different values for ${{dT_c^{\prime}} \over {dP}}$ between
105 and 160 K/GPa were obtained in different studies of C$_{60}$ with
most of the data located around 110 K/GPa \cite{samara,Kriza,samara2}.
Bulk modulus of the FCC phase was reported to be 16.4 GPa
\cite{Bashkin} that is intermediate between other values
\cite{Duclos,Fisher}. For low-symmetry phase the bulk modulus was
found to be 11.6 GPa \cite{LundinSSC}. It gives the discontinuity of
isothermal compressibility $\Delta \beta_T\approx$
2.5$\times$10$^{-2}$ GPa$^{-1}$ and implies $\Delta \alpha_{_V}
\approx$ 22.5$\times$10$^{-5}$ K$^{-1}$ with $\Delta {\cal C}_P
\approx$ 0.23 kJ/(mol~K) through the Keesom-Ehrenfest relationship
Eq.(\ref{keeserenf}). This $\Delta \alpha_{_V}$ is in qualitative
accord with results of the diffraction study \cite{HeineyPRB} where
$\alpha_{_V}$ for FCC phase was found to be 6.1$\times$10$^{-5}$
K$^{-1}$ and definite precursor phenomena were observed in
low-temperature phase with average $\alpha_{_V} \approx$
21$\pm$3$\times$10$^{-5}$ K$^{-1}$. Our value of $\Delta {\cal C}_P$
appears to be twice as much as experimental value
\cite{ACcalorimetry}.

The dimensionless volume effect $\Delta \epsilon_0$ was found around
-0.01 \cite{David,HeineyPRB} and it gives the absolute hysteresis
\begin{eqnarray*}
T_0 - T_c = {B^2 \over {4 \alpha^{\prime} C^{\prime}}} =
- \frac{9}{4} {\Delta \epsilon_0 \over {\Delta\beta_T}}
{dT_c \over dP} \approx 99 \ {\rm K} \ .
\end{eqnarray*}
Taking the highest reported transition temperature 261.5 K
\cite{Dworkin} as $T_0^{\prime}$ we get $T_c^{\prime} \approx$ 162.5 K
and $T_{\ast}^{\prime}(P=0) \approx$ 251 K. This is in very good
agreement with DSC experiments \cite{GriveiPRB} where two peaks around
250 K and 150 K were found in ${\cal C}_P$ of C$_{60}$ upon cooling
and only sharp peak at 260 K has been found during heating. Peak at
150 K can be attributed to the critical order parameter fluctuations
in supercooled high-temperature phase near $T_c$ whereas peak around
250 K corresponds to the onset of the transition accompanied by
increase in ${\cal C}_P$ of low-symmetry phase near $T_0$. This peak
increases in superheated SC phase during heating cycle. Sharp peak in
the sound velocity and attenuation was found near 160 K in C$_{60}$
single-crystal \cite{Shi} and the anomaly in the X-ray diffraction
peak intensity was also reported around 155 K \cite{xrayanom}.

\begin{figure}
\epsfxsize 5.75in
\centerline{\epsffile{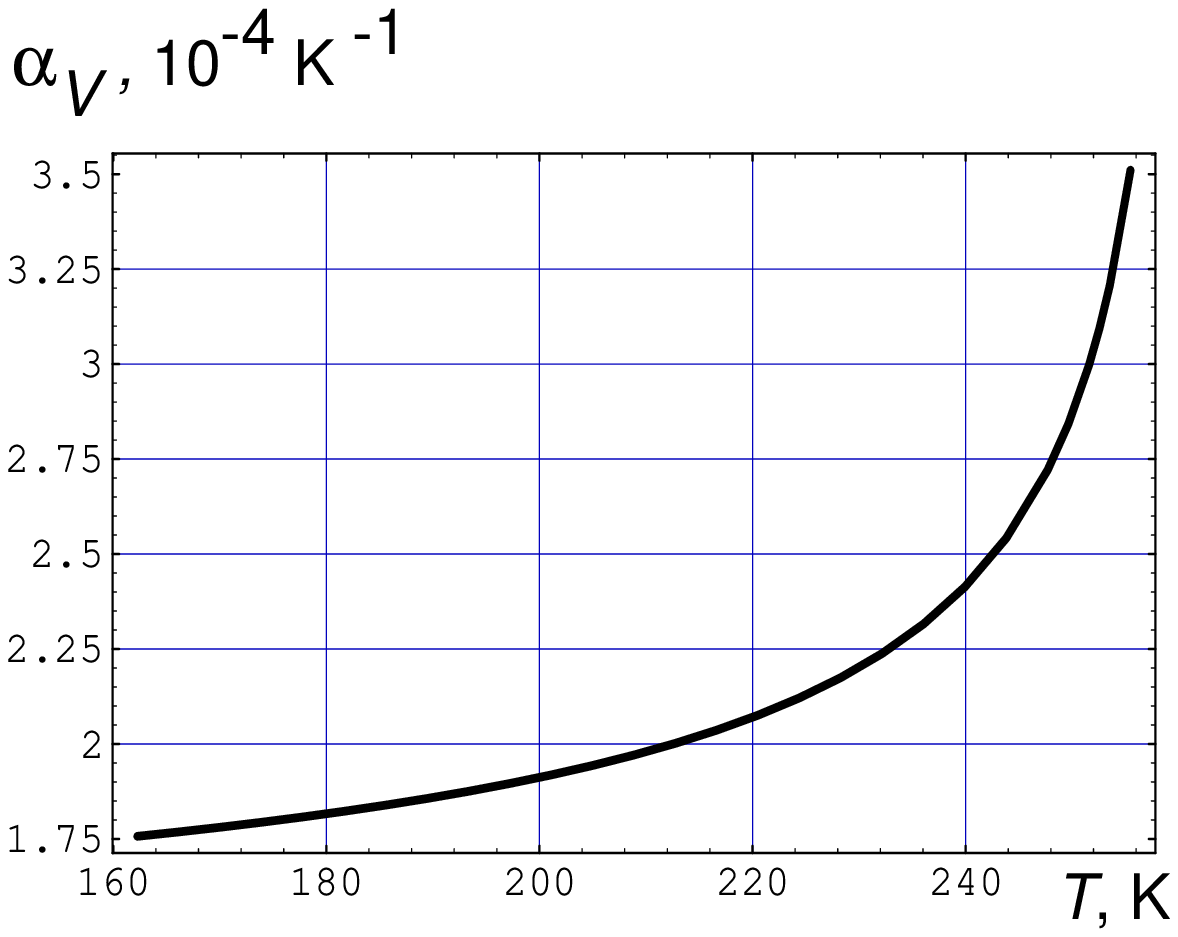}}
\caption{Temperature dependence of the volume thermal expansion
	coefficient in low-symmetry phase.}
        \label{expansion}
\end{figure}

\begin{figure}
\epsfxsize 5.75in
\centerline{\epsffile{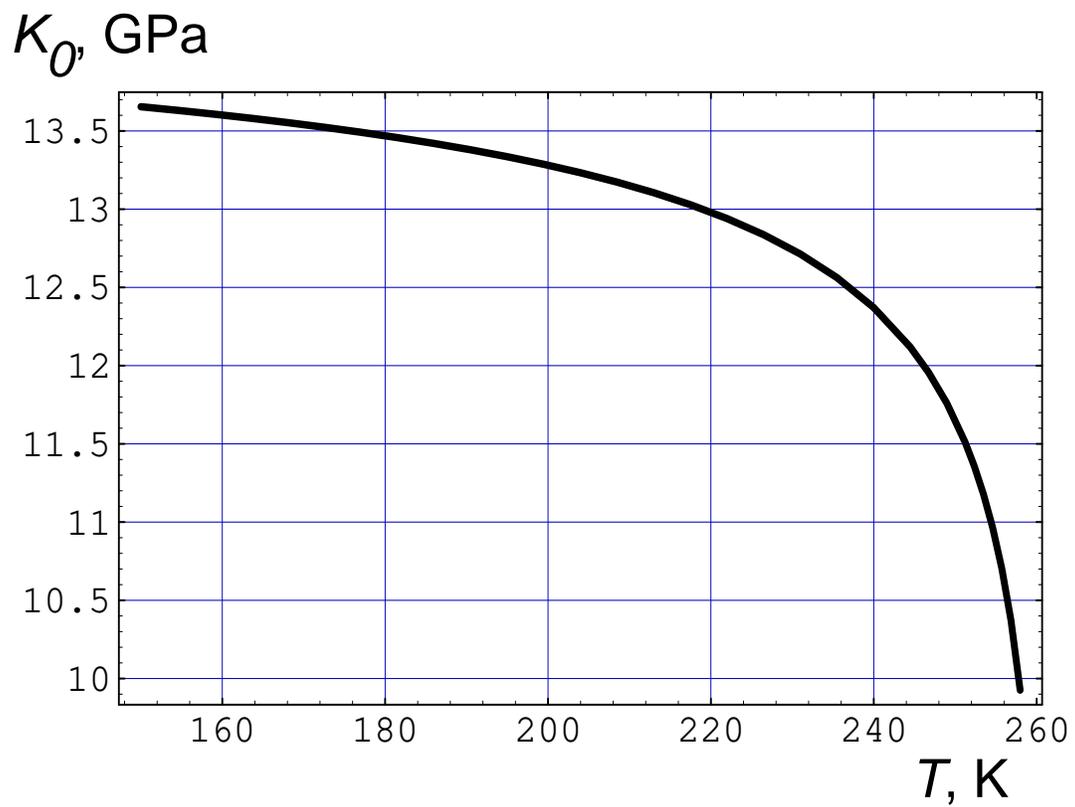}}
\caption{Bulk modulus of SC phase {\em vs} temperature.}
         \label{bulktemp}
\end{figure}

\begin{figure}
\epsfxsize 5.75in
\centerline{\epsffile{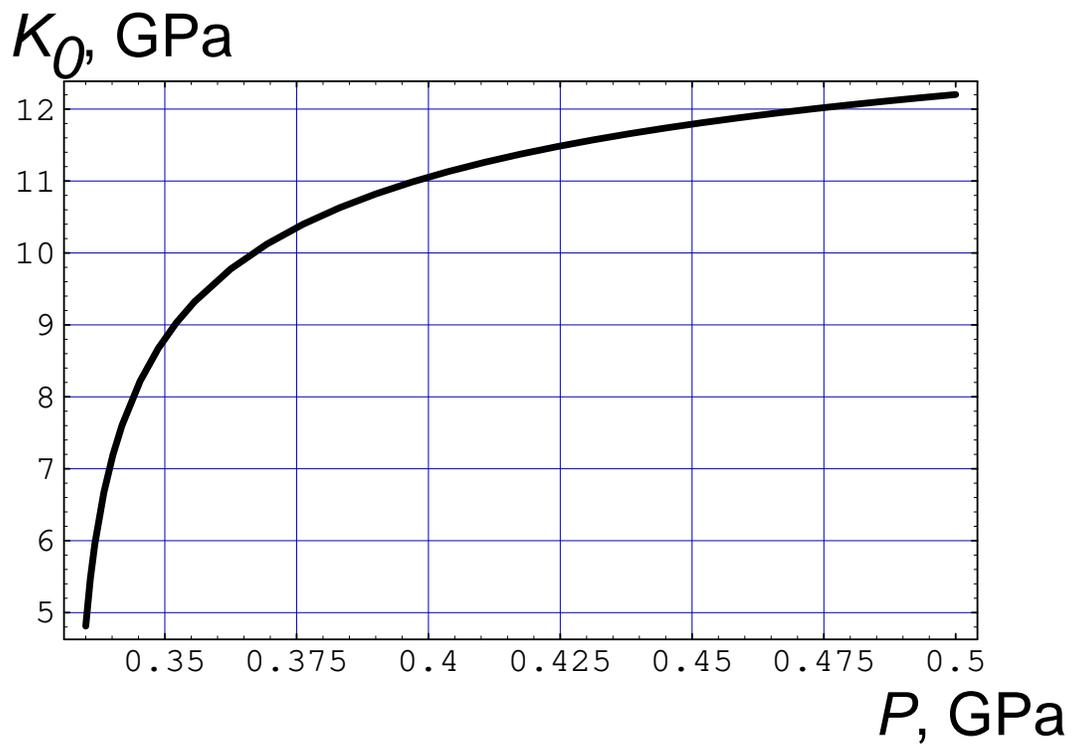}}
\caption{Pressure dependence of bulk modulus in SC phase
        for room temperature 25$^o$C.}
        \label{bulkmod}
\end{figure}

The temperature dependence of $\alpha_{_V}$ in low-temperature phase
plotted according to Eq.(\ref{expcoeff1}) is shown in
Fig.\ref{expansion}, where qualitative agreement with the experimental
data \cite{Gugenberger} can be seen. Similar expression appears for
specific heat and isothermal compressibility, and the temperature
dependence of bulk modulus in SC phase shown in Fig.\ref{bulktemp} can
be obtained accordingly. Pressure acts through the change of
$T_c^{\prime}$ by Eq.(\ref{tempres}) and the pressure dependence of
bulk modulus in the low-symmetry SC phase
\begin{eqnarray}
\label{modpres}
K(T,P) & = & K_0 \Biggl( 1 + {2 D_0^2 \over {C^{\prime} K_0}}\,
\biggl( 1 + \nonumber \\ & & \ \ \ \left( {T_0 - T \over {T_0 - T_c}} +
{8 D_0 C^{\prime} P \over {K_0 B^2}} \right)^{-{1 \over 2}}
\biggr) \Biggr)^{-1} ,
\end{eqnarray}
is plotted in Fig.\ref{bulkmod} for room temperature.
Eq.(\ref{modpres}) is essentially different from the Murnaghan
equation $K(P) = K(0) + K^{\prime} P$ used in some experimental
studies to describe measured data. The assumption of linear dependence
of bulk modulus on pressure has led to conclusion about negative jump
in compressibility upon transition from FCC to low-symmetry SC phase
at $P = 0$ \cite{Lundin94} even though in the same study positive
compressibility jump was found at 293 K for observed pressure-induced
transition around 0.2 GPa. As the transition discontinuities
(\ref{discont}) do not depend on applied hydrostatic pressure and
volume effect has been indeed found independent on $P$ in that
experiment, one can suggest that increase in $K$ is a result of linear
extrapolation of obtained data outside the region of the stability of
SC phase at $P=0$. The Murnaghan equation was found inadequate in
recent pressure experiment \cite{Ludwig} as well, where bulk modulus
of SC phase at 170 K was found to be 14.2 GPa, that is in qualitative
agreement with our $K(P)$ curve in Fig.(\ref{bulkmod}).

Thus, thermoelastic singularities associated with the orientational
ordering transition in crystalline fullerene C$_{60}$ can be
reproduced in the Landau theory through temperature and pressure
dependence of the Gibbs free energy in SC phase and coupling of
elastic strain with order parameter. The Keesom-Ehrenfest relations
(\ref{keeserenf}) for second derivatives of the Gibbs free energy
appear to be satisfied, hence, the energy is continuous around
transition and can, thus, be naturally written in the Ginzburg-Landau
form due to symmetry breaking.

Usual belief is that the singularities around phase transition are due
to effect of critical fluctuations and for many transitions of the
'order-disorder' kind the fluctuations are known to be strong. Thus,
agreement obtained in the present approach within the Landau theory
which ignore the fluctuations implies question why they do not make
significant contribution in this case. Possible answer is that the
transition temperature $T_{\ast}^{\prime}$ is located far from
critical temperature $T_c$, where high-temperature phase becomes
unstable and considerable fluctuations arise. One can suppose that the
coupling of the order parameter with strain suppress inhomogeneous
fluctuations because elastic forces have long-range behavior and the
strain energy associated with order parameter fluctuations is large.
In order to make quantitative estimation the experimental information
concerning correlation length for orientational fluctuations is
needed.

We can evaluate, however, the homogeneous order parameter fluctuations
in the present model. In order the weakly-first-order transition to be
properly identified and clearly separated from the background of
fluctuations in high-temperature phase, the discontinuity of the order
parameter should be greater than its thermal fluctuations
(\ref{fluct}). It gives us criterion for the third-order coefficient:
\begin{equation}
B \ \ \gg \ \  \left( C^3 k_B T_c \right)^{1 \over 4} .
        \label{Bcond}
\end{equation}
We have
\begin{displaymath}
{B^4 \over {{C^{\prime}}^3}} =
\frac{81}{2}\,{(\Delta \epsilon_0)^2 \over {\Delta \beta_T}}
\approx 69.6 \ {\rm kJ/mol} \ ,
\end{displaymath}
that is much larger than the energy of thermal fluctuations
$k_B T_{\ast}^{\prime} \approx$ 2.08 kJ/mol and, hence, the homogeneous
order parameter fluctuations are very weak near $T_{\ast}^{\prime}$.

To conclude, we have analyzed the thermoelastic phenomena around the
orientational ordering transition in crystalline fullerene C$_{60}$ in
the frame of the mean-field Landau theory. The transition
discontinuities of thermal expansion coefficient and isothermal
compressibility originate from the temperature and pressure dependence
of the Gibbs free energy in low-symmetry phase through coupling of
elastic homogeneous strain with order parameter. The temperature range
of the coexistence of FCC and SC phase appears to be very large - from
260 K down to critical temperature $T_c \approx$ 150 K and it can
explain experimentally observed anomalies around this point.

\acknowledgments

Discussions with P. Dolinar, V. Kaganer and K.H. Michel were very
helpful. The work would have not been done without hospitality of
Prof. J. Goldak at Carleton University.

\end{document}